\pdfoutput=1 
\documentclass{JINST}
\usepackage{graphicx}
\usepackage{amsmath,amsthm}
\usepackage{url}
\usepackage[mediumspace,mediumqspace,squaren]{SIunits}
\usepackage{subfigure}
\usepackage{makerobust}
\MakeRobustCommand\subref

\title{\bf Study of glass properties as electrode for RPC}

\author{K. Raveendrababu$^{a,c}$, P. K. Behera$^{a}$, B. Satyanarayana$^{b}$ and Jafar Sadiq$^{a}$ \\
\llap{$^{a}$}Physics Department, Indian Institute of Technology Madras, Chennai 600036, India \\
\llap{$^{b}$}Department of High Energy Physics, Tata Institute of Fundamental Research, Mumbai 400005, India \\
\llap{$^{c}$}Physical Science Division, Homi Bhabha National Institute, Anushaktinagar, Mumbai 400094, India \\
E-mail: \email{prafulla.behera@gmail.com}}

\abstract{
Operation and performance of the Resistive Plate Chambers (RPCs) mostly depend on the quality and characteristics of the electrode 
materials. The India-based Neutrino Observatory collaboration has chosen glass RPCs as the active detector elements for its Iron Calorimeter 
detector and is going to deploy RPCs in an unprecedented scale. Therefore, it is imperative that we study the electrode material aspects in 
detail. We report here, systematic characterization studies on the glasses from two manufacturers. RPC detectors were built using these glasses 
and performances of the same were compared with their material properties.
}


\begin{document}

\maketitle                 

\section{Introduction}
\label{Sec:Introduction}
{
The India-based Neutrino Observatory (INO) collaboration has proposed to build a 50 kiloton magnetized Iron Calorimeter (ICAL) to detect
the atmospheric neutrinos and antineutrinos separately. Main aims of the INO-ICAL experiment are to make precision measurements on the 
atmospheric neutrino oscillation parameters and address the fundamental issue of neutrino mass hierarchy. The INO-ICAL will consist of three 
modules of 16 $\times$ 16 $\times$ 14.5 m$^3$. Each module will have a stack of 151 horizontal layers of 5.6 cm thick iron plates interleaved 
with 4 cm gaps to house the active detectors. The experiment is expected to run for more than 10 years to record statistically sufficient 
neutrino oscillation data~\cite{1}. Layout of the proposed INO-ICAL detector, comprising of three modules is shown in Figure~\ref{Fig:1}.

The INO collaboration has chosen 2 $\times$ 2 m$^2$ size glass Resistive Plate Chambers (RPCs) as the active detector elements and is going to 
deploy 28,800 of them in the ICAL detector. It is proposed to use glasses from Asahi and/or Saint-Gobain for producing RPCs. Therefore, 
we performed systematic material property studies on the glass samples from these two manufacturers and compared the performances of RPCs 
built using these glasses. 

\begin{figure}[ht]
        \centering
         \includegraphics[width=0.7\textwidth]{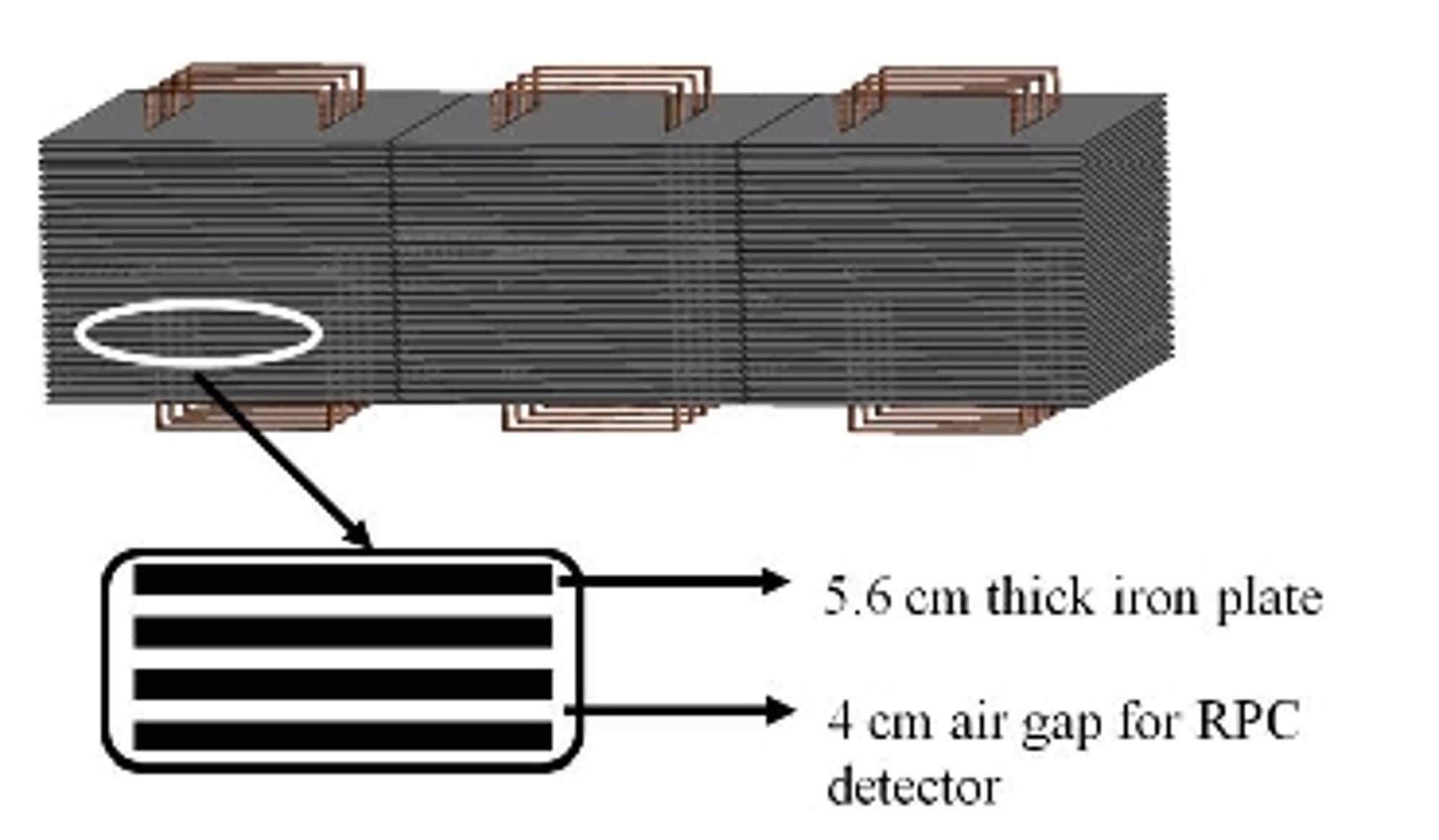}
          \caption {Layout of the proposed INO-ICAL detector.}
\label{Fig:1}
\end{figure}
}

\subsection{Resistive plate chamber} 
\label{Sec:Resistive plate chamber}
\vskip 0.2cm
{Resistive Plate Chambers (RPCs) are gaseous detectors, which work on the ionization principle. They are simple to construct in large sizes, 
offer two-dimensional readout, provide good efficiency (>90\%) and time resolution ($\sim$1 ns). Therefore, they are used as the trigger and/or 
timing detectors in many high-energy physics experiments~\cite{2}. The basic construction scheme of a single gap RPC is shown in 
Figure~\ref{Fig:2}.

\begin{figure}[ht]
        \centering
         \includegraphics[width=0.81\textwidth]{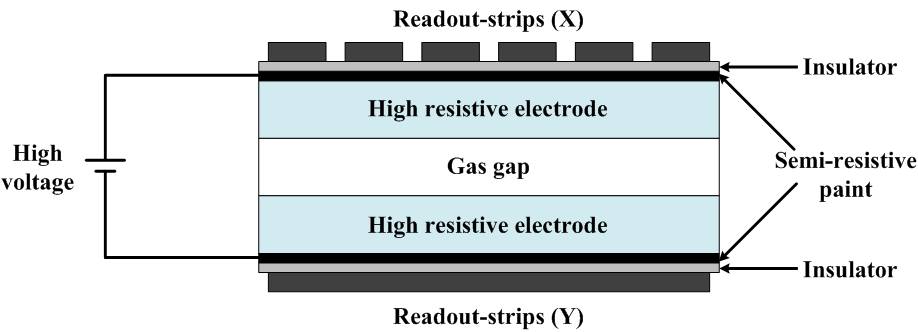}
          \caption {Basic construction schematic of a Resistive Plate Chamber.}
\label{Fig:2}
\end{figure}
}

\section{Float glass characterization}

\subsection{Surface roughness}
\label{Sec:Surface roughness}
{
The inner surface of the electrode that is facing the gas volume should be as smooth as possible, so that the field emission of electrons 
from cathode can be minimized~\cite{3}. These emitted electrons increase dark current and singles rate of RPC and thereby deteriorating its 
performance. Therefore, the surface roughness of glass samples from Asahi and Saint-Gobain was measured using BRUKER ContourGT Optical 
Microscope. The microscope resolution is <0.01 nm. The surface roughness measurements are shown in Figure~\ref{Fig:3}. The RMS surface 
roughness (Rq) values of the glasses are quoted below the image of each scan. Both the glasses showed identical surface roughness. 

\begin{figure}[ht]
        \centering
         \includegraphics[width=0.7\textwidth]{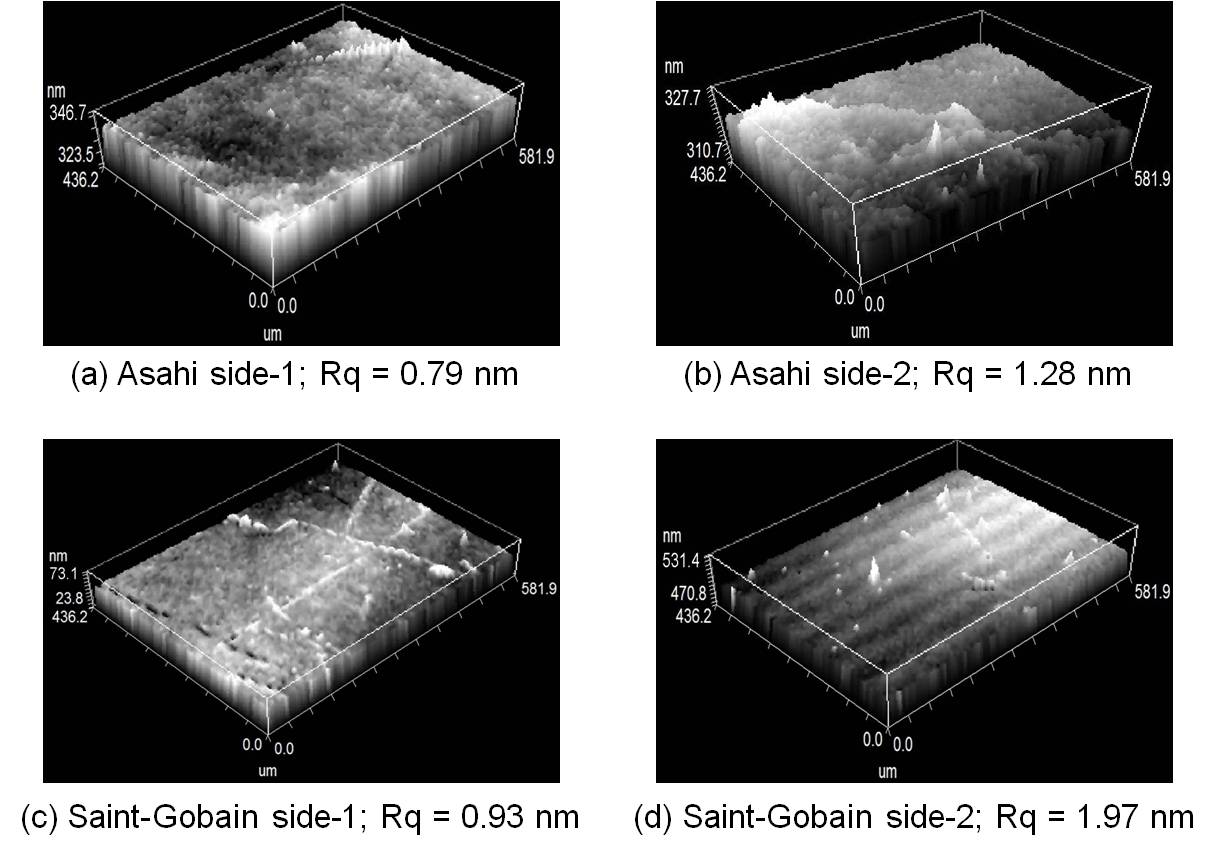}
          \caption {Surface roughness measurements of the glasses using BRUKER ContourGT Optical Microscope. Size of each scan is 
                    582 $\mu$m $\times$ 436 $\mu$m.}
\label{Fig:3}
\end{figure}

}

\subsection{Elemental composition}
\label{Sec:Elemental composition}
{
Elemental compositions of the glass samples were measured using Energy Dispersive X-ray technique equipped with FEI Quanta 200 Scanning 
Electron Microscope. The results in fractional atomic percentages are summarized in Table~\ref{Tab:1}. We observed that Asahi glass is 
showing larger Na component ($\sim$ 2\%) compared to that of Saint-Gobain glass.

\begin{table}[ht]
\caption{Summary of various elemental compositions of glass samples in fractional atomic percentages (\%).}
\begin{center}
\vskip -0.6cm
\resizebox{\textwidth}{!}{
\begin{tabular}{|l|c|c|c|c|c|c|}
\hline
{\bf Element} & \multicolumn{3}{c|}{\bf Asahi (\%)} & \multicolumn{3}{c|}{\bf Saint-Gobain (\%)} \\ 
\cline{2-7}
&\bf Sample-1 &\bf Sample-2 &\bf Average &\bf Sample-1 &\bf Sample-2 &\bf Average \\
\hline
Oxygen (O) & 70.93 & 68.65 & 69.79 & ~71.03 & 71.38 & 71.21 \\

Silicon (Si) & 15.43 & 16.69 & 16.06 & 16.7 & 16.86 & 16.78 \\

Sodium (Na) & 10.46 & 10.71 & 10.59 & ~8.7 & ~8.29 & 8.5 \\

Magnesium (Mg) & ~2.21 & ~2.51 & ~2.36 & ~~2.17 & ~2.31 & ~2.24 \\

Calcium (Ca) & ~0.48 & ~0.82 & ~0.65 & ~~0.85 & 0.6 & ~0.73 \\

Aluminium (Al) & ~0.48 & ~0.56 & ~0.52 & ~~0.46 & ~0.52 & ~0.49 \\

Iron (Fe) & ~0.02 & ~0.05 & ~0.04 & ~~0.08 & ~0.04 & ~0.06 \\
\hline
\end{tabular}
}
\label{Tab:1} 
\end{center} 
\end{table}  
}

\subsection{Relative permittivity}
\label{Sec:Relative permittivity}
{
Relative permittivities of glass samples were measured using 'Novocontrol Broadband Dielectric/Impedance Spectrometer' at 
$20 \ ^{\circ}$C. We observed that the relative permittivity of Asahi glass is larger compared to that of Saint-Gobain glass as shown in 
Figure~\ref{Fig:4}. This may be related to larger Na component in Asahi glass as shown in Table~\ref{Tab:1}. In glass, the larger Na component 
leads to larger Na$^+$O$^-$ ionic characteristic bonds. These bonds are more easily polarizable in an applied electric field, which leads to 
larger electric permittivity~\cite{4}. 

\begin{figure}[ht]
        \centering
         \includegraphics[width=0.6 \textwidth]{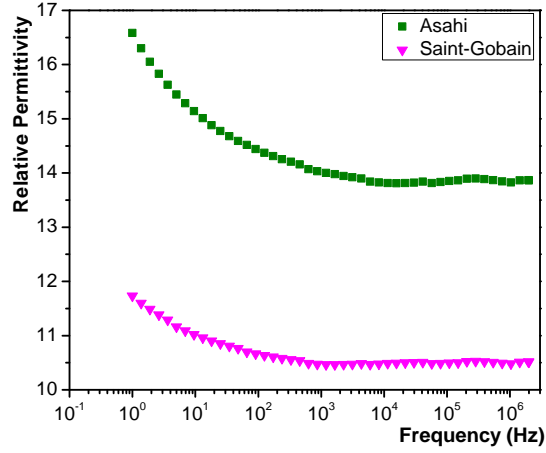}
\vskip -0.5cm
          \caption{Relative permittivity measurements of Asahi and Saint-Gobain glasses using Novocontrol Broadband 
                   Dielectric/Impedance Spectrometer.}
\label{Fig:4}
\end{figure}
}

\section{RPC performance studies}
\label{Sec:RPC performance studies}

\subsection{Experimental setup}
\label{Sec:Experimental setup}

RPCs of 30 $\times$ 30 cm$^2$ size were built using 3 mm thick float glass plates from Asahi and Saint-Gobain manufacturers. The outer surfaces 
of the glass plates were coated with specially developed conductive graphite paint for the ICAL RPCs, which facilitates applying high voltage 
across the electrodes~\cite{5,6}. The surface resistance of the electrodes was uniform ($\sim$1 M$\Omega$/ \hspace{-5mm}$\qed$). 
Readout-strips of 2.8 cm wide, with 0.2 cm gap between the consecutive strips, were orthogonally mounted on the external surfaces of the RPCs. 
A layer of mylar sheet was inserted between the electrode and the readout-strip. The gas gaps made out of Asahi glass and Saint-Gobain glass 
were named as A-RPCs and S-RPCs, respectively. A gas mixture of C$_{2}$H$_{2}$F$_{4}$/iso-C$_{4}$H$_{10}$/SF$_{6}$ = 95/4.5/0.5 was flown 
through the RPCs with a total flow rate of 10 SCCM and operated in the avalanche mode. All the RPCs were operated under identical environmental 
conditions.

\begin{figure}[ht]
        \centering
         \includegraphics[width=0.9 \textwidth]{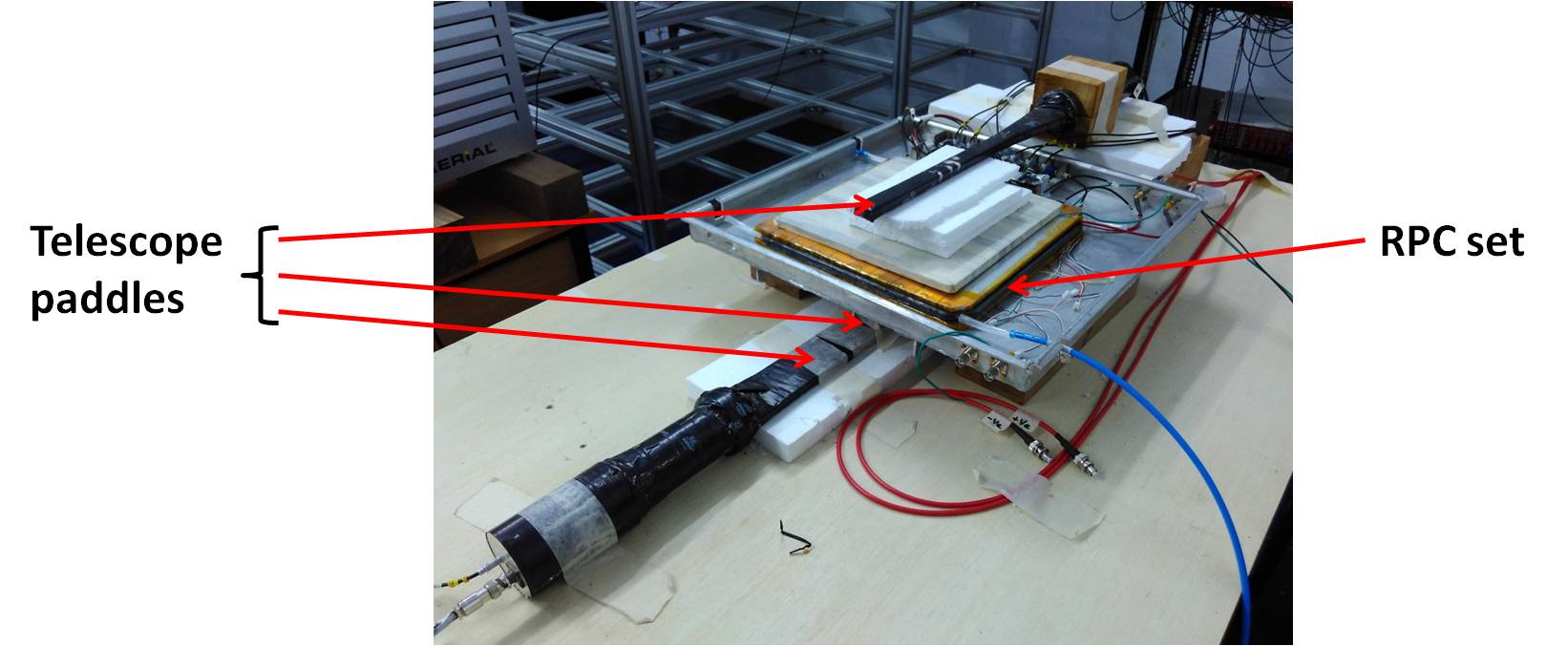}
          \caption{Experimental setup developed for characterizing the 30 $\times$ 30 cm$^2$ size RPCs.}
\label{Fig:5}
\end{figure}

A cosmic ray muon telescope was set up with three plastic scintillation paddles to get a 3-fold coincidence. The dimensions of scintillation 
paddles in length $\times$ width $\times$ thickness are 30 $\times$ 2 $\times$ 1 cm$^3$ (top), 30 $\times$ 3 $\times$ 1 cm$^3$ (middle), and 
30 $\times$ 5 $\times$ 1 cm$^3$ (bottom). The RPCs were stacked between top and middle scintillation paddles. The experimental setup 
developed to test the RPCs is shown in Figure~\ref{Fig:5}.

\subsection{Results}
\label{Sec:Results}
{
\noindent {\bf Voltage-current characteristics:} \\
The voltage-current characteristics of the RPCs were measured using C.A.E.N Mod. N471A, 2 channel HV Power Supply. The current resolution 
of the module is 1 nA. The currents drawn by the RPCs as a function of applied voltage are shown in Figure~\ref{Fig:6}. A-RPCs were found 
to draw lower bias currents compared to S-RPCs. This is due to larger relative permittivity of Asahi glass compared to Saint-Gobain as shown in 
Figure~\ref{Fig:4}. In an applied electric field, the larger relative permittivity material shows comparatively smaller leakage currents. 
 
\begin{figure}[ht]
        \centering
         \includegraphics[width=0.6 \textwidth]{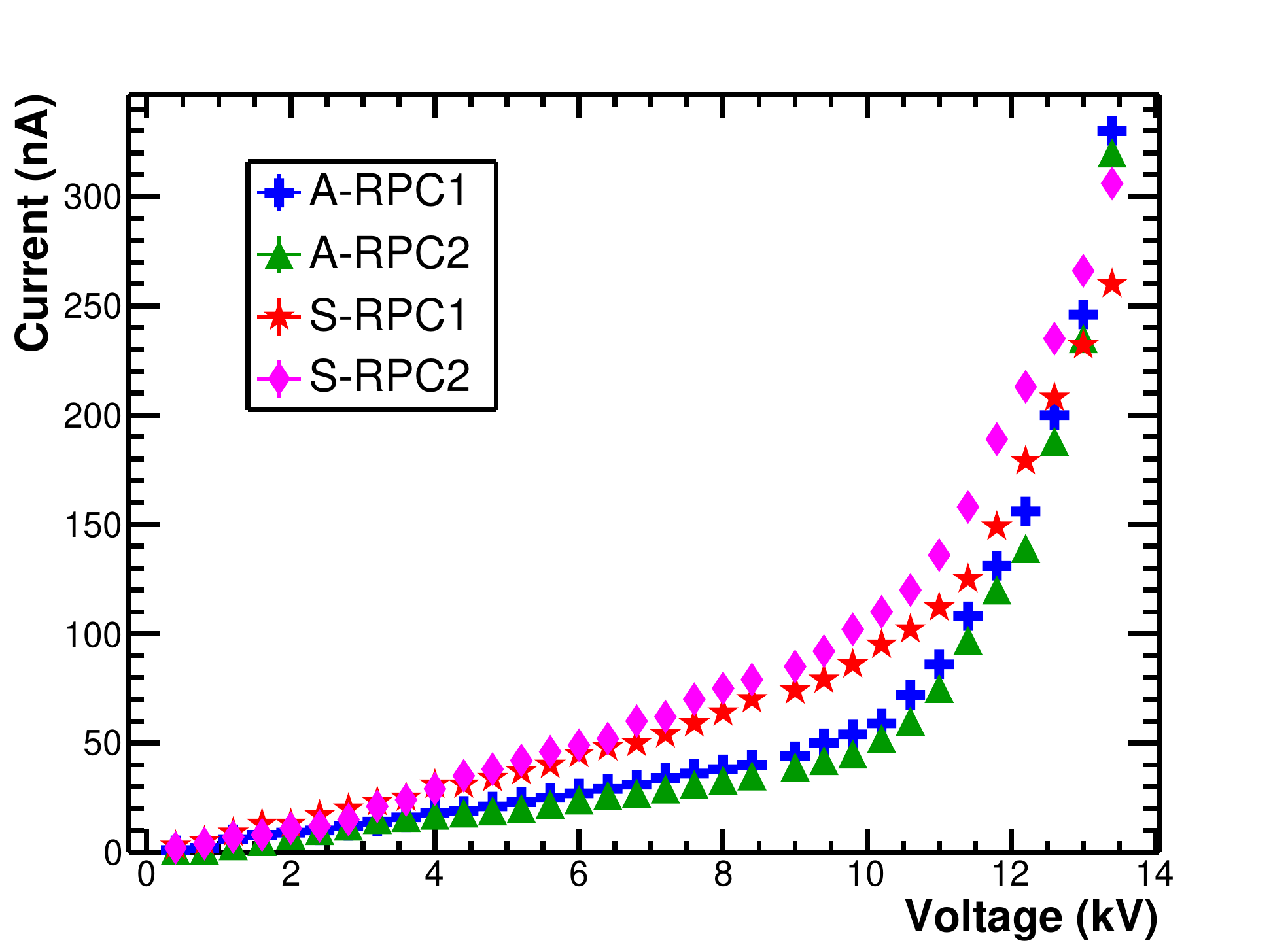}
          \caption{Currents drawn by the RPCs as a function of applied voltage.}
\label{Fig:6}
\end{figure}

\vskip 0.5cm

\noindent {\bf Efficiency studies:} \\
The normal vector component of the electric field displacement is continuous at the electrode/gas interface of RPC. This boundary condition 
is expressed as:
\begin{equation}
{\bf D} = \epsilon_{p} {\bf {E}_{p}} = \epsilon_{g} {\bf {E}_{g}},
\label{Eq:1} 
\end{equation}

\noindent
where, $\epsilon_{p}$ and $\epsilon_{g}$ are the permittivities, and ${\bf {E}_{p}}$ and ${\bf {E}_{g}}$ are the electric fields of the 
electrode plate and the gas gap, respectively~\cite{7}. Equation~\ref{Eq:1} indicates that the RPCs made out of electrodes with 
larger relative permittivity can be operated at lower bias voltages.

Efficiencies of the RPCs were measured using cosmic-ray muons. The experimental setup is shown in Figure~\ref{Fig:5}. We used C.A.E.N Mod. V830 
Scaler for these measurements. The efficiencies of the RPC detectors were calculated using the formula:

\begin{equation}
 Efficiency = {\frac{4\text{-}fold \ (4F) \ rate}{3\text{-}fold \ (3F) \ rate}} \times 100 \%,
\label{Eq:2}
\end{equation}

\noindent
where, $3F$ is the telescope trigger signal, which is generated by time coincidence of three scintillator paddles, $4F$ is 
the coincidence signal of the RPC under test and the telescope trigger signal. 

\vskip 0.5cm

The efficiencies of RPCs as a function of applied high voltage are shown in Figure~\ref{Fig:7}. All four RPCs showed more than 95\% efficiency 
on the plateau. It is observed that the knee of efficiency plateau of A-RPCs starts at 10.8 kV, where as that of S-RPCs starts at 12.0 kV. 
Therefore, A-RPCs can be operated at 1.2 kV lower bias voltage in comparison to S-RPCs. As already mentioned above, this is due to larger 
relative permittivity of Asahi glass and hence these efficiency results are consistent with Equation~\ref{Eq:1}.

\begin{figure}[ht]
        \centering
         \includegraphics[width=0.6 \textwidth]{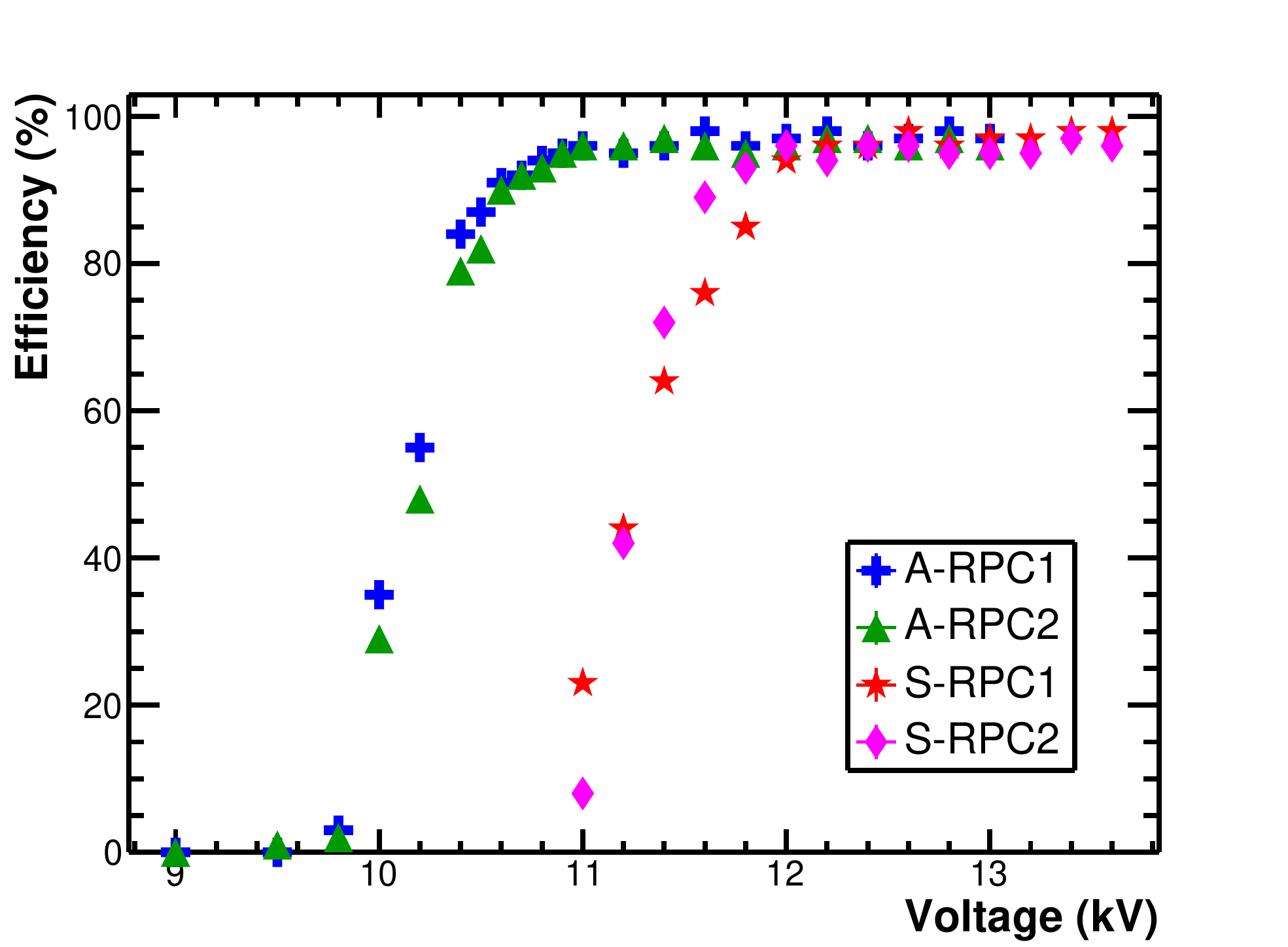}
          \caption{Efficiencies of the RPCs as a function of high voltage.}
\label{Fig:7}
\end{figure}
}

\section{Conclusions}
\label{Sec:Conclusions}
{
Systematic characterization studies were performed on the float glass samples from Asahi and Saint-Gobain. They showed identical surface
roughness. From the elemental composition measurements, we observed that Na component of Asahi glass is larger ($\sim$ 2\%) compared to that 
of Saint-Gobain glass. We conclude that this is the reason for higher relative permittivity of Asahi glass. RPC detectors were built using 
these glasses and their performances were studied. Asahi RPCs were found to draw lower bias currents compared to Saint-Gobain RPCs. The knee of 
efficiency plateau of Asahi RPCs and Saint-Gobain RPCs started at 10.8 kV and 12.0 kV, respectively. Therefore, our study indicates that the 
RPCs made from Asahi glass will be better suited for the INO-ICAL experiment. 
}      

\section{Acknowledgements}
\label{Sec:Acknowledgements}
{
This work was supported by the Department of Atomic Energy (DAE), and the Department of Science and Technology (DST), Government of India. 
The authors would like to gratefully acknowledge the help of their colleagues V. Asgolkar, S. Chavan, R. R. Shinde, V. M. Datar, N. K. Mondal 
at TIFR, Mumbai, and V. Janarthanam at IIT Madras. 
}

\end{document}